\documentclass{epl}

\title{Supersonic Discrete Kink-Solitons and Sinusoidal Patterns with 
``Magic" Wavenumber in Anharmonic Lattices}
\author{Yuriy A. Kosevich \inst{1} Ramaz Khomeriki \inst{2} \and Stefano Ruffo \inst{3,4}}
\institute{
\inst{1} Instituto de Investigacion en Comunicacion Optica, Universidad Autonoma de San 
\\ Luis Potosi, Alvaro Obregon 64, 78000 San Luis Potosi, S.L.P., Mexico \\
\inst{2} Department of Physics, Tbilisi State University, 
3 Cha\-vchava\-dze avenue \\ Tbilisi 380028, Republic of Georgia \\
\inst{3} Dipartimento di Energetica ``Sergio Stecco" and CSDC Universit\`a di Firenze, 
\\Via S. Marta, 3, I-50139 Firenze, Italy \\
\inst{4} INFM and INFN Firenze, Italy 
}


\begin{document}

\maketitle

\begin{abstract}
The sharp pulse method is applied to Fermi-Pasta-Ulam (FPU) and Lennard-Jones 
(LJ) anharmonic lattices. Numerical simulations reveal the presence of high 
energy strongly localized ``discrete'' kink-solitons (DK), which move with 
supersonic velocities that are proportional to kink amplitudes. For small 
amplitudes, the DK's of the FPU lattice reduce to the well-known ``continuous''
kink-soliton solutions of the modified Korteweg-de Vries equation. For high 
amplitudes, we obtain a consistent description of these DK's in terms of 
approximate solutions of the lattice equations that are obtained by restricting 
to a bounded support in space exact solutions with sinusoidal pattern 
characterized by the ``magic'' wavenumber $k=2\pi/3$. Relative displacement 
patterns, velocity versus amplitude, dispersion relation and exponential tails 
found in numerical simulations are shown to agree very well with analytical 
predictions, for both FPU and LJ lattices.
\end{abstract} 

\vspace{-0.6cm}

\section{Introduction}

The interest in studying moving nonlinear localized excitations (solitons,
kink-solitons, etc.) is mainly motivated by the fact that they can be 
related to transport properties. In this respect, one 
dimensional lattices of anharmonic oscillators can serve as a testing ground for
nonlinear transport. 
In this Letter we present numerical and analytical studies of the dynamical 
properties of strongly localized excitations of anharmonic lattices 
with two-body interparticle potentials~\cite{fermi}. Invariance under the 
symmetry transformation 
that shifts the positions of all the oscillators by the same amount 
relates anharmonic lattices with interparticle couplings to a 
wide class of systems which share 
such continuous symmetry, e.g. quasi-one dimensional easy plane ferromagnets 
and antiferromagnets, ferrimagnetic systems with spiral structures 
and even quantum Hall double layer pseudo-ferromagnets~\cite{magnets}.
The spontaneous breakdown of this translational symmetry leads to the apperance of 
gapless Goldstone modes, which itself generates kink-soliton solutions in the 
presence of nonlinearity. 
Anomalous transport properties~\cite{anomalous} could appear for all systems in 
this class. It is extremely important to assess whether such properties 
could be related to the existence of moving localized excitations.
In the weakly nonlinear continuum limit and for gapless dispersion relations one 
can prove~\cite{japan} that all localized solutions can be described by the 
Korteweg-de Vries (KdV) or modified KdV (mKdV) 
equations~\cite{zabusky,yuriy1,yuriy2,poggi0,ramaz2}. 
On the lattice, small-amplitude solutions can be represented as KdV or mKdV solitons 
in terms of relative displacements,  while they are kinks for absolute displacements. 
The energy of such objects is proportional to the kink amplitude and thus  
lattice discreteness has to be taken into account when considering high 
energy (i.e. strongly localized) objects. Since in this limit one cannot 
use the ordinary continuum approximation, Rosenau \cite{rosenau1} has 
recently proposed a theoretical approach that takes explicitly into account 
lattice discreteness. Applying this approach to the 
Fermi-Pasta-Ulam (FPU) lattice~\cite{fermi}, he derived, in the limit
of strong nonlinearity, a different 
continuum equation, which supports a new kind of localized solution that 
he called {\it compacton}~\cite{rosenau}.

On the other hand, the existence of approximate kink-like solutions of
the FPU lattice with characteristic sinusoidal displacement pattern with
the ``magic" wavenumber $k=2\pi/3$ has been predicted on the basis of a direct 
analysis of the discrete lattice equations \cite{kosevich}.
This ansatz relies on the existence of exact extended propagating
nonlinear sinusoidal waves having this particular wavenumber~\cite{kosevich,poggi}. 
Such patterns are also extremely successful in modelling stationary and 
slowly-moving discrete breathers in the FPU lattice~\cite{kosevich,lepri1,yuriy3}.

The numerical study of moving strongly localized excitations of anharmonic 
lattices has been so far limited to discrete breathers (DB)~\cite{flach} that
move with small velocity. 
Two types of numerical methods have been commonly used to study the time 
evolution of DB's: {\it i)} exact~\cite{flach,aubry} or approximate 
\cite{sandusky,takeno} DB's are put initially onto the lattice and {\it ii)} 
modulational 
instability of zone-boundary and large-wavenumber modes is 
induced~\cite{lepri1,yuriy3,burlakov,peyrard,ruffo,ullmann,mirnov}. 
The first method allows one to investigate single
moving localized objects, but it assumes the knowledge of exact or approximate 
solutions. By using the second method, a number of
localized modes is produced: they move, collide and merge, and it is 
therefore hard to investigate each single moving object. 
On the other hand, no mathematical 
proof of existence of moving DB's has been obtained (see \cite{sepulchre} for
a discussion of this issue and a possible way out).

In this Letter we show a new numerical method to produce strongly localized
objects. This method is inspired by the one used in real experiments
to generate optical solitons in fibers as well as 
magnetization envelope solitons in Yttrium Iron Garnet film 
waveguides~\cite{yttrium}. A sharp pulse 
is applied at one end of an unperturbed sample. Then, the system itself chooses 
the stable 
propagating localized mode and the only thing that is left is
to follow its motion.  
We have first performed these numerical experiments
for an FPU chain with quadratic and quartic
interactions between neighbors, which describes purely
transverse vibrational excitations.
For large amplitudes, we observe supersonic strongly
localized discrete kink-solitons (DK) in which just three oscillators are
excited. 
They show very low energy loss. We have repeated these experiments for more 
realistic Lennard-Jones (LJ) interparticle potentials, also revealing the existence 
of DK's. 

Fast moving kinks have been observed in the Frenkel-Kontorova (FK) 
model~\cite{braun,savin}. Moving kinks appear also in translationally invariant 
lattices with Hertz potentials (see e.g. Ref.~\cite{rosas} and references therein), 
but in this case there is no attractive force and the localized wave dynamics is 
different from that of the FPU lattice.

\section{Sharp Pulse Method}

The equations of motion of the FPU-$\beta$ lattice~\cite{fermi} are
\begin{equation}
\ddot u_n=u_{n+1}+u_{n-1}-2u_n+(u_{n+1}-u_n)^3+(u_{n-1}-u_n)^3, 
\label{2}
\end{equation}
where $u_n$ is the transverse displacement of the $n$-th oscillator from its equilibrium 
position in dimensionless units. 

In the numerical experiment we oscillate the first particle of the chain in the 
following manner:
\begin{eqnarray}
u_1&=&A_0\cdot sin\bigl[\omega_0(t_0-t)\bigr]  \qquad \mbox{if} \qquad 0<t\leq 
t_0 \qquad \mbox{and} \nonumber \\ &&
u_1=0 \qquad \mbox{if} \qquad t>t_0,
\label{1} 
\end{eqnarray}
while the right end $u_{N}$ is pinned. After applying the above pulse, 
we pin both ends of the lattice. 

The numerical experiment shows the existence of kinks that propagate with 
large and slightly different velocities. Throughout this paper we will use 
two different representations of the spatial profile: either we use absolute 
lattice displacements $u_n$ or we show relative displacements 
$v_n=u_{n+1}-u_n$. Hence the kinks are shown as true steps in the first 
representation (see Fig. 1a) or as sharp peaks in the second representation 
(see Fig. 1b). 

As time evolves, the plateaus between the DK's widen but the shape of the 
kinks does not change. During the collisions, the kink-solitons cross each 
other without any appreciable energy loss (this is not shown here for FPU 
lattices but later on for LJ lattices, see the right graph in Fig. 4).  
%%%%%%%%%%%%%%%%%%%%%%%%%%%%%%
\begin{figure}[b]
%h=here, t=top, b=bottom, p=separate figure page
\begin{center}\leavevmode
\includegraphics[height=5cm]{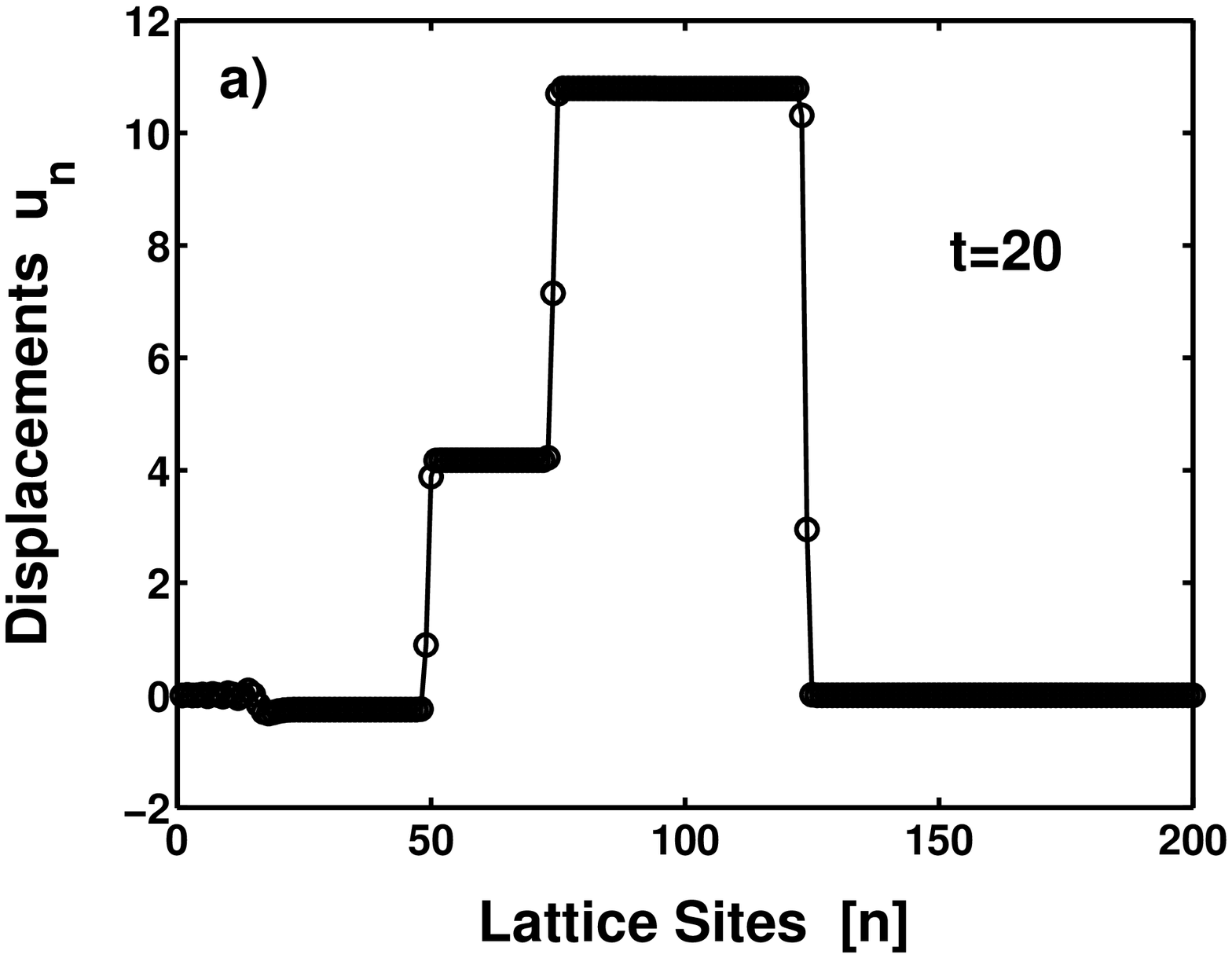}
\includegraphics[height=5cm]{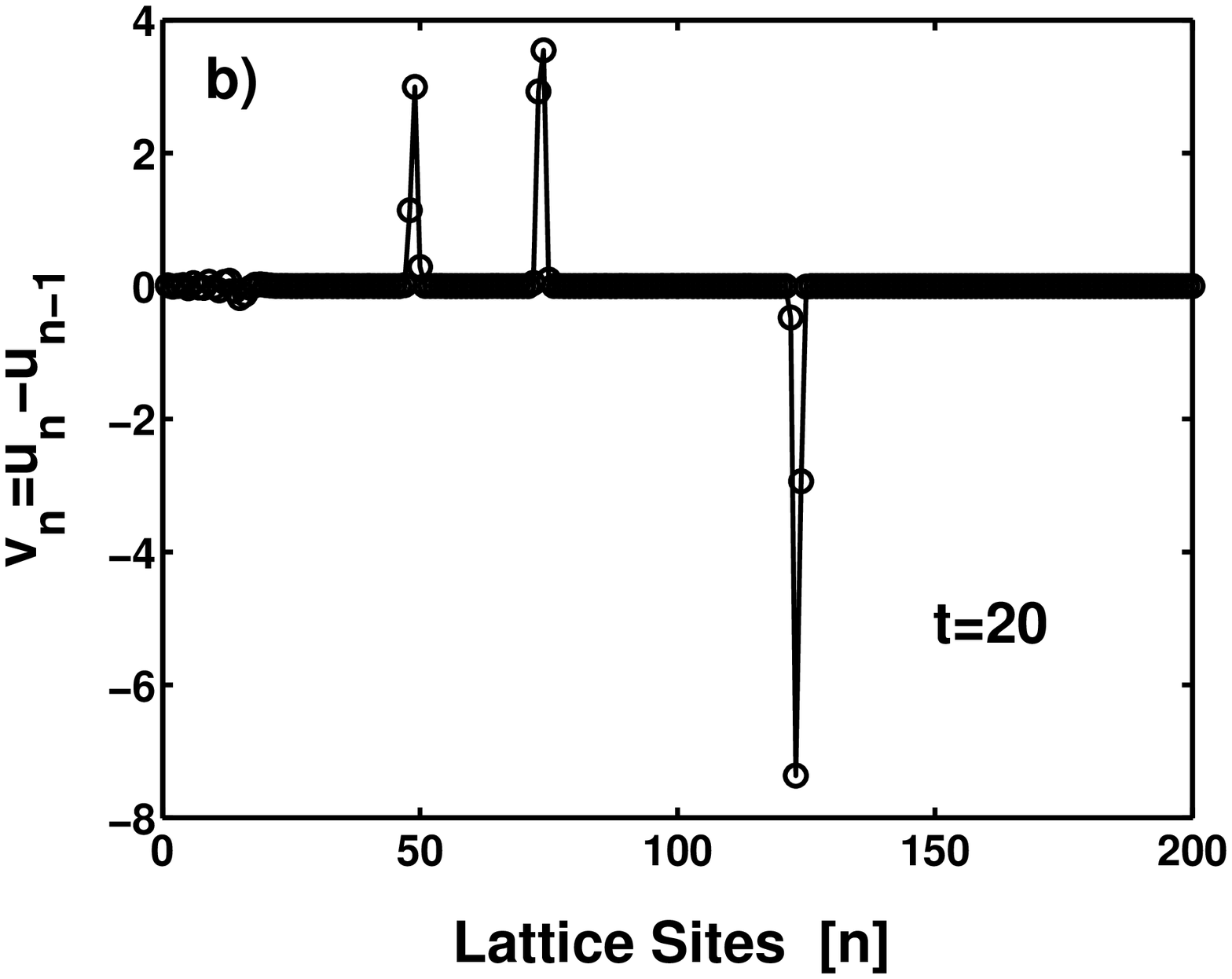}
\end{center}
\caption{\label{fig1} a) Absolute $u_n$ and b) relative $v_n=u_{n+1}-u_{n}$ 
displacements versus the lattice site $n$ at time $t=20$ in dimensionless units.
In the $u$-space we observe a sequence of kinks with different amplitudes, 
while in the $v$-space the localized objects have solitonic form of different 
height. This simulation is performed with $A_0=11$, $\omega_0=\sqrt{2}$, $t_0=1$ 
and $N=200$.}
\end{figure}
%%%%%%%%%%%%%%%%%%%%%%%%%%%%%%%%%%%%%%%%%%%%

The existence of these localized objects is possible because of the peculiar
symmetry $u_n\rightarrow u_n+const$ of the interparticle potential. Systems with
such a symmetry have an infinitely degenerate ground state and its spontaneous
breakdown leads to the appearance of gapless Goldstone modes that determine the
formation of kinks.

Since the potential energy of the FPU chain is only a function of the relative 
displacements $v_n=u_{n+1}-u_{n}$, these are the variables to be considered for 
the appropriate physical picture. In these new variables the equations of motion 
(\ref{2}) can be exactly rewritten as 
\begin{equation}
\ddot v_n=v_{n+1}+v_{n-1}-2v_n+v_{n+1}^3+v_{n-1}^3-2v_n^3. \label{4}
\end{equation}
Eq. (\ref{4}) leads, in the continuum approximation and selecting one direction
for the wave motion, to the modified Korteweg-de Vries 
(mKdV) equation if one keeps only the first terms in the two small parameters 
$1/N$ and 
amplitude~\cite{zabusky}. The mKdV equation supports exact soliton 
solutions. Indeed, for small pulse amplitudes we could observe mKdV solitons. 
However, as the driving amplitude $A_0$ of the pulse is increased we observe the 
DK's of Fig. 1. Therefore, in terms of relative displacements, the objects that 
we 
observe in our simulations can be considered as discrete extension of the 
solutions 
of the mKdV equation. 
The existence of fast moving objects with amplitude 
independent short localization length was first considered in 
Ref.~\cite{kosevich}. In the remaining of this Letter, we develop a
theoretical approach that, making use of this result allows us to describe 
the main physical properties of the DK's we observe in our simulations.

\section{Approximate Analytical Description}

Exact extended sinusoidal wave solutions of the FPU (and other) lattices have been 
recently found by several authors~\cite{kosevich,poggi,chechin}. One of such solutions 
has the ``magic" $k=2\pi/3$ wavenumber and can be written in terms of relative 
displacements
as $v_n=A\cos(kn-\omega t)$. It has been proposed that such exact solution can acquire 
a compact support and maintain its validity as approximate solution~\cite{kosevich}. 
For instance, in the case of stationary and slowly-moving discrete 
breathers in the FPU-$\beta$ 
lattice, half a period truncations of the sinusoidal envelope of the displacement
patterns with $k_{e}=\pi - k=\pi/3$ wavenumber 
well reproduce large amplitude localized oscillations~\cite{kosevich,lepri1,yuriy3}.
Following the same strategy, we here propose a similar truncation that turns out 
to describe very well our supersonic DK's:
\begin{eqnarray}
v_n&=&\pm\frac{A}{2}\left[1+\cos\right(\frac{2\pi}{3}n-\omega t\left)\right] 
\qquad \mbox{if} \qquad -\pi<\frac{2\pi}{3}n- \omega t<\pi; 
\label{111}
\end{eqnarray} 
and $v_n=0$ otherwise. In the large amplitude limit we can neglect 
short exponential tails in the sinusoidal pattern (\ref{111}) (see, however, Eq. (6) 
below). Substituting this ansatz into Eq~(\ref{4}) 
and performing the rotating wave approximation  \cite{rwa} for the 
second harmonic, we obtain the following relations for the frequency 
$\omega$ and velocity $V$ as the functions  of the amplitude $A$ of the DK's.
\begin{equation}
\omega=\sqrt{3+(45/16)A^2}; \qquad V=\omega/(2\pi/3)
=3\sqrt{3+(45/16)A^2}/(2\pi). \label{112}
\end{equation}

%%%%%%%%%%%%%%%%%%%%%%%%%%%%%%
\begin{figure}[t]
%h=here, t=top, b=bottom, p=separate figure page
\begin{center}\leavevmode
\includegraphics[height=5cm]{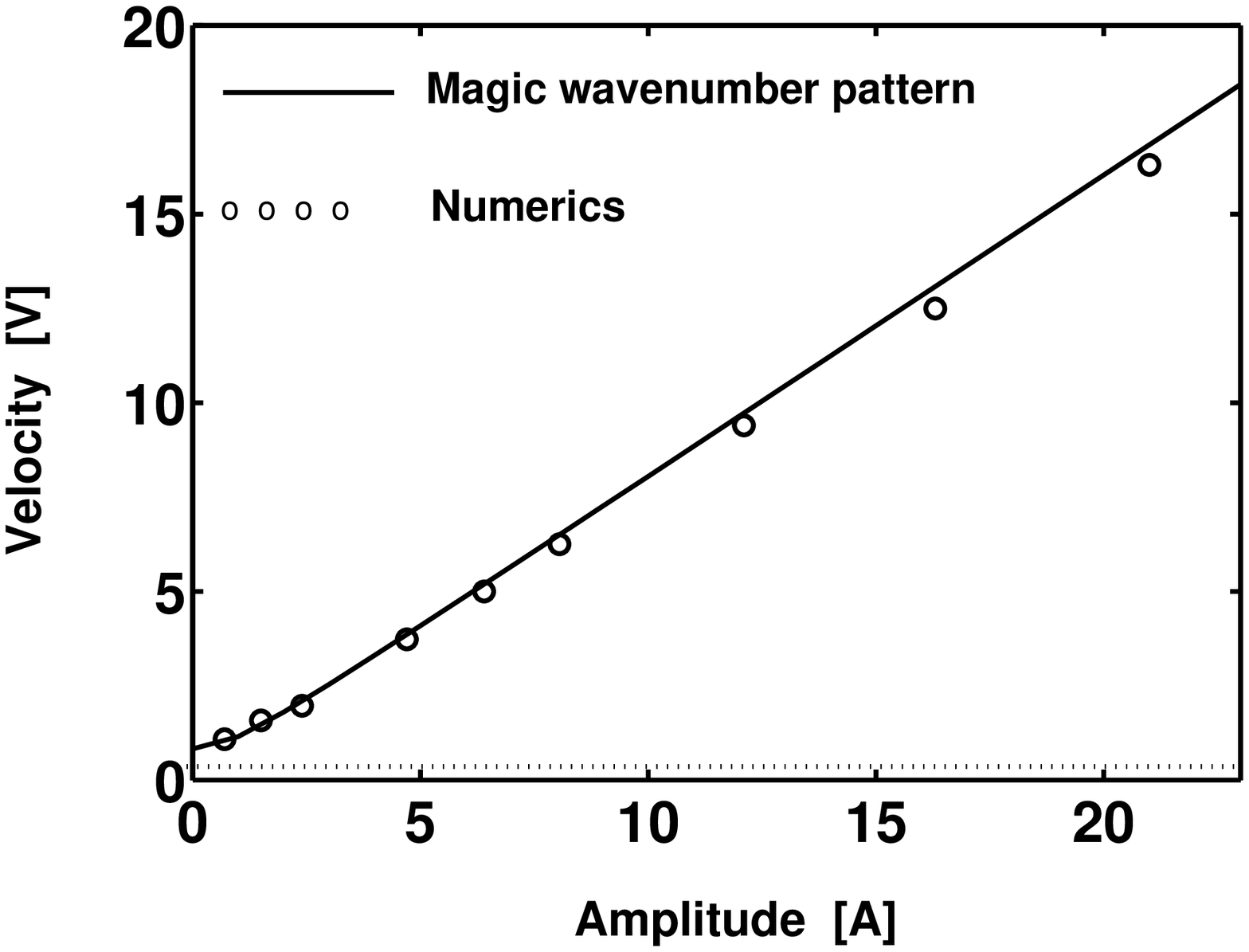}
\includegraphics[height=5cm]{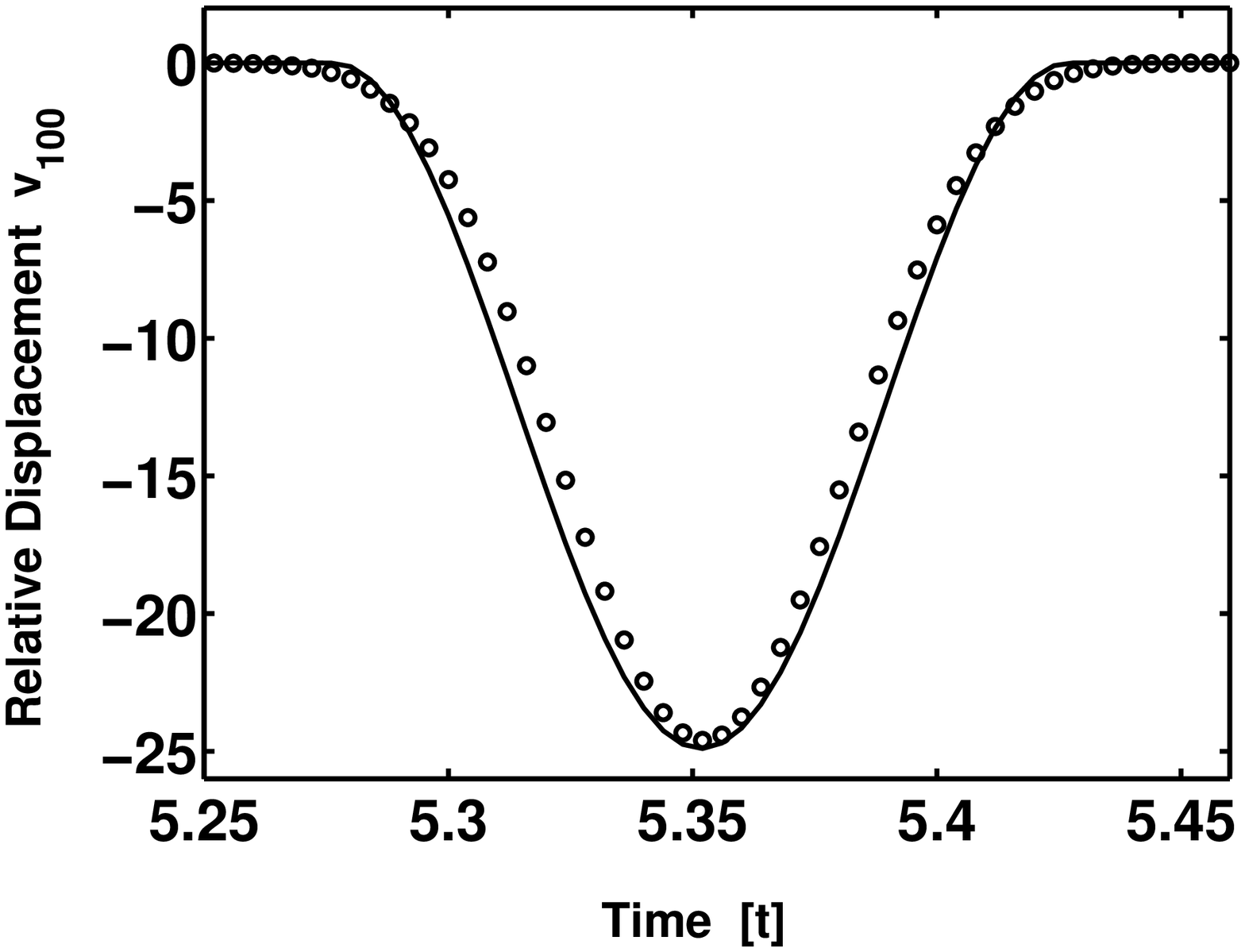}
\end{center}
\caption{\label{fig2} Left graph: Velocity of the localized discrete 
kink-solitons versus amplitude of relative displacements. Open circles represent 
the result of numerical experiments. The solid line corresponds to 
the analytical formula (\ref{112}), which follows from the ``magic" wavenumber
solution. The dotted horizontal line 
indicates the maximal group velocity of linear waves, which equals 1. 
Right graph: Relative displacements of a given oscillator versus time. Open 
circles are the results of numerical measurements, the 
solid line corresponds to formula (\ref{111}).}
\end{figure}
%%%%%%%%%%%%%%%%%%%%%%%%%%%%%%%%%%%%%%%%%%%%

First of all, we can observe that the analytical solution of formula (\ref{111}) 
is consistent with the numerical observation that three 
lattice sites are always excited (see Fig.1). Besides that, we have 
numerically checked the dependence of velocity on amplitude 
(see left graph in Fig. 2) and the time dependence of the relative
displacement of a given oscillator (see right graph in Fig. 2).
Our theoretical expression gives a very good agreement with all numerical data.
Furthermore, the right graph of Fig. 2 shows that the relative displacement 
patterns of both the measured and the analytical solution has everywhere a
smooth envelope. Therefore, we can conclude that the compacton with sharp 
cosine-like support and with 
non-smooth relative displacement envelope, described in Refs.
~\cite{rosenau1,rosenau}, doesn't compare well with our numerical results.
We'll come back to this and related issues in a forthcoming 
publication~\cite{newus}.

In relation to this latter comment, it is important to remark that our DK's have
a localized bulk sided by exponential tails (see Fig. 3a). The decay length 
$\Lambda$ is a function of the velocity, as shown in Fig. 3b. This dependence
can be derived analytically representing the relative displacements in the tails as
 $v_n\sim \exp{\left[\pm(n-Vt)/\Lambda\right]}$ and substituting this ansatz into 
equations (\ref{4}). Considering the limit $n-Vt\gg \Lambda$, one gets the 
relation~\cite{kosevich}:
\begin{equation}
V^2=4\Lambda^2\sinh^2(1/2\Lambda). \label{tail}
\end{equation}
As follows from Eq.~(6), the decay length $\Lambda$ is real positive only for
supersonic excitations with $V>1$ and it diverges for $V=1$. Again, the agreement with
numerical data, shown in Fig. 3b, is very good.

To check the stability of the approximate solution (\ref{111}), we directly put it on 
the unperturbed lattice, imposing to the kink-soliton the initial velocity given by 
formula (\ref{112}). In several simulations we observe the long-living motion 
of a slightly modified kink-soliton. Since in our
experiments the boundaries are pinned, we observe edge reflection. Both at edge 
reflections and during collisions, the kink-solitons do not significantly 
change their amplitudes and shapes.
%%%%%%%%%%%%%%%%%%%%%%%%%%%%%%
%%%%%%%%%%%%%%%%%%%%%%%%%%%%%%
\begin{figure}[t]
%h=here, t=top, b=bottom, p=separate figure page
\begin{center}\leavevmode
\includegraphics[height=4.5cm]{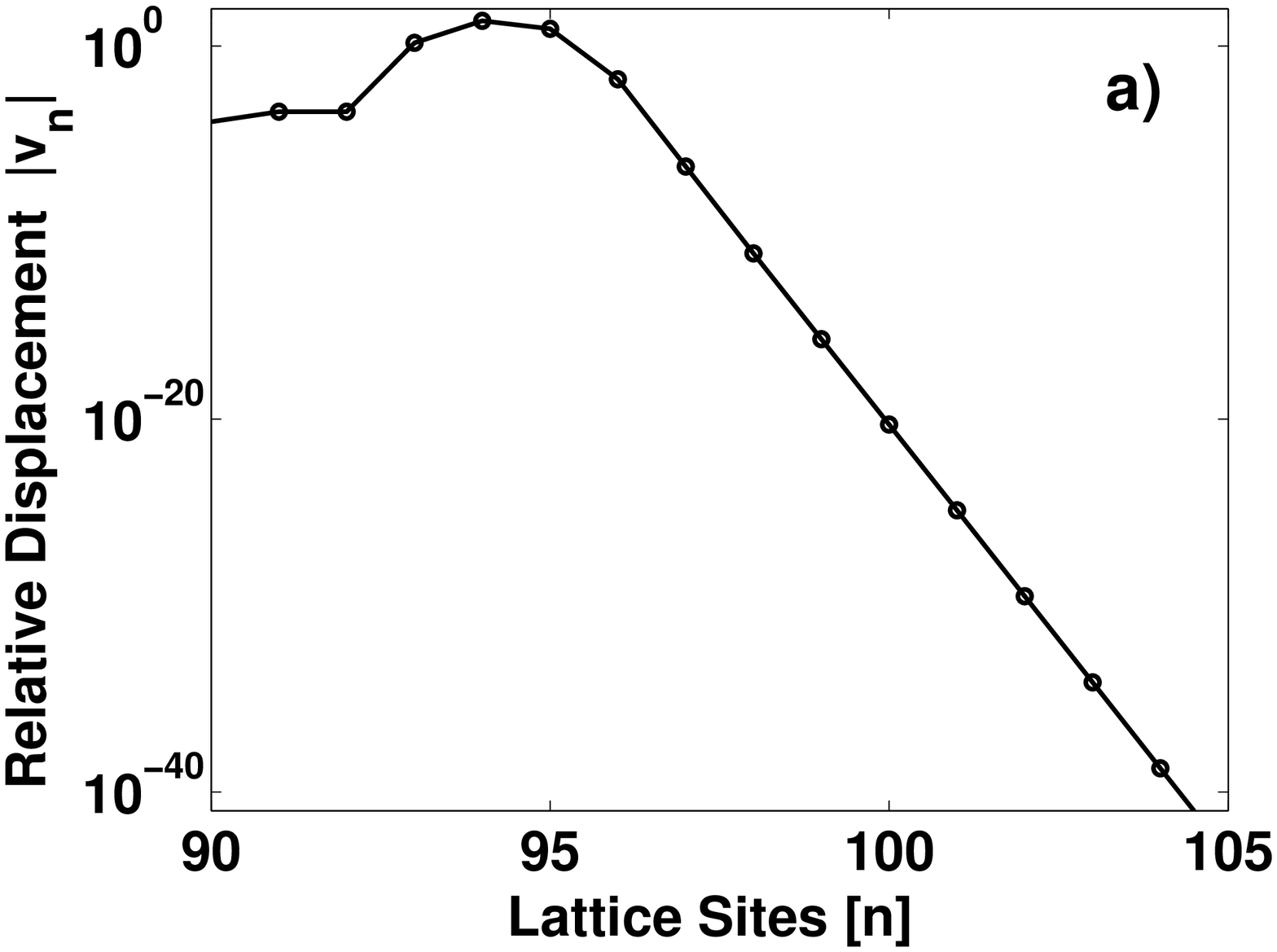}
\includegraphics[height=4.5cm]{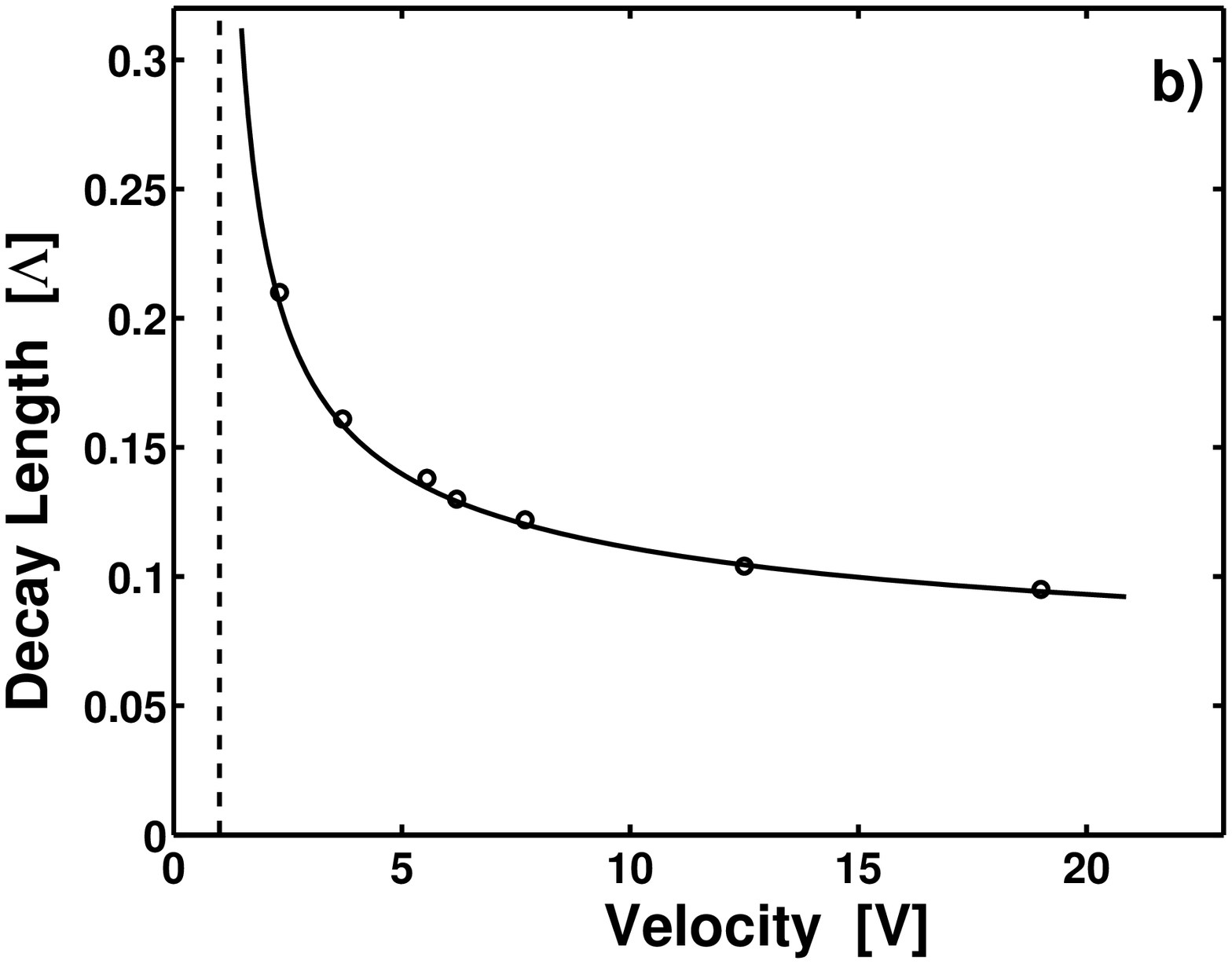}
\end{center}
\caption{\label{fig3} a) Logarithmic plot of the relative displacement 
pattern of a single DK which moves from the left to the right. We use 
the right exponential decay, which is not affected  by  radiation, to
fit the decay length $\Lambda$. b) Dependence of the decay length, $\Lambda$, 
upon the velocity of the DK. The solid line is obtained from formula (\ref{tail}) and 
the points are results of numerical experiments. The dashed line shows the 
velocity, $V=1$, at which the decay length diverges.}
\end{figure}
%%%%%%%%%%%%%%%%%%%%%%%%%%%%%%%%%%%%%%%%%%%%

\section{Realistic Potentials}

In order to check the generality of our results we have performed a similar 
analytical and numerical study of the Lennard-Jones (LJ) interparticle potential 
$V(r)=4e\left[\left(\frac{s}{r}\right)^{12}-\left(\frac{s}{r}\right)^6\right]$, 
where $r$ is a distance between neighboring particles, $e$ and $s$ are constants. 
After the appropriate rescalings of space and time: 
$r\rightarrow (2)^{1/6}s\cdot r$ and ~~$t\rightarrow t\sqrt{m/12e}$,  
we obtain the following equations of motion for the relative displacements:
\begin{equation}
\frac{d^2v_n}{dt^2}=\frac{2}{(1+v_n)^{13}}- \frac{1}{(1+v_{n-1})^{13}} -\frac{1}{(1+v_{n+1})^{13}}- 
\frac{2}{(1+v_n)^{7}}+ \frac{1}{(1+v_{n-1})^{7}} +\frac{1}{(1+v_{n+1})^{7}}. \label{lennard}
\end{equation}

The dispersion relation for linear waves can be derived by substituting the pattern
$v_n=Acos(kn-\omega t)$ in Eq.~(\ref{lennard}). In the $A\rightarrow 0$ limit
we get the value of maximal group velocity of linear waves 
$v_{max}=\sqrt{6}$ 
(see dashed line in the left panel of Fig. 4).

Seeking for strongly localized solutions, we substitute again the ``magic" 
wavenumber sinusoidal pattern (\ref{111}) into the equations of motion (\ref{lennard}). 
We choose the minus sign in the relative displacement pattern (\ref{111}),
which corresponds to the conditions under which we observe supersonic
kinks in LJ lattices. Then, we expand the r.h.s. of the equations of motion in 
powers of $(A/2)\cos(2\pi n/3-\omega t)$, keeping terms
up to fifth order. To treat the powers of cosines we again perform 
the rotating wave approximation~ \cite{page}.
We don't report here the formula that we obtain for the velocity-amplitude
relation to save space, but its graph is plotted in the left panel in Fig. 4 
as a solid line. 
The agreement between analytics and numerics is quite satisfactory,
confirming that the ``magic" wavenumber ansatz works also for LJ
potentials. We have also investigated numerically the DK's interaction process. 
As seen from the right panel of Fig. 4, kink-solitons retain their shapes and 
velocities after interaction. The effect of the interaction manifests itself
only in 
tiny trajectory shifts, as it is expected from the weakly nonlinear continuum
approximation (see e.g. \cite{ramaz2}).

%%%%%%%%%%%%%%%%%%%%%%%%%%%%%%
\begin{figure}[t]
%h=here, t=top, b=bottom, p=separate figure page
\begin{center}\leavevmode
\includegraphics[height=4.5cm]{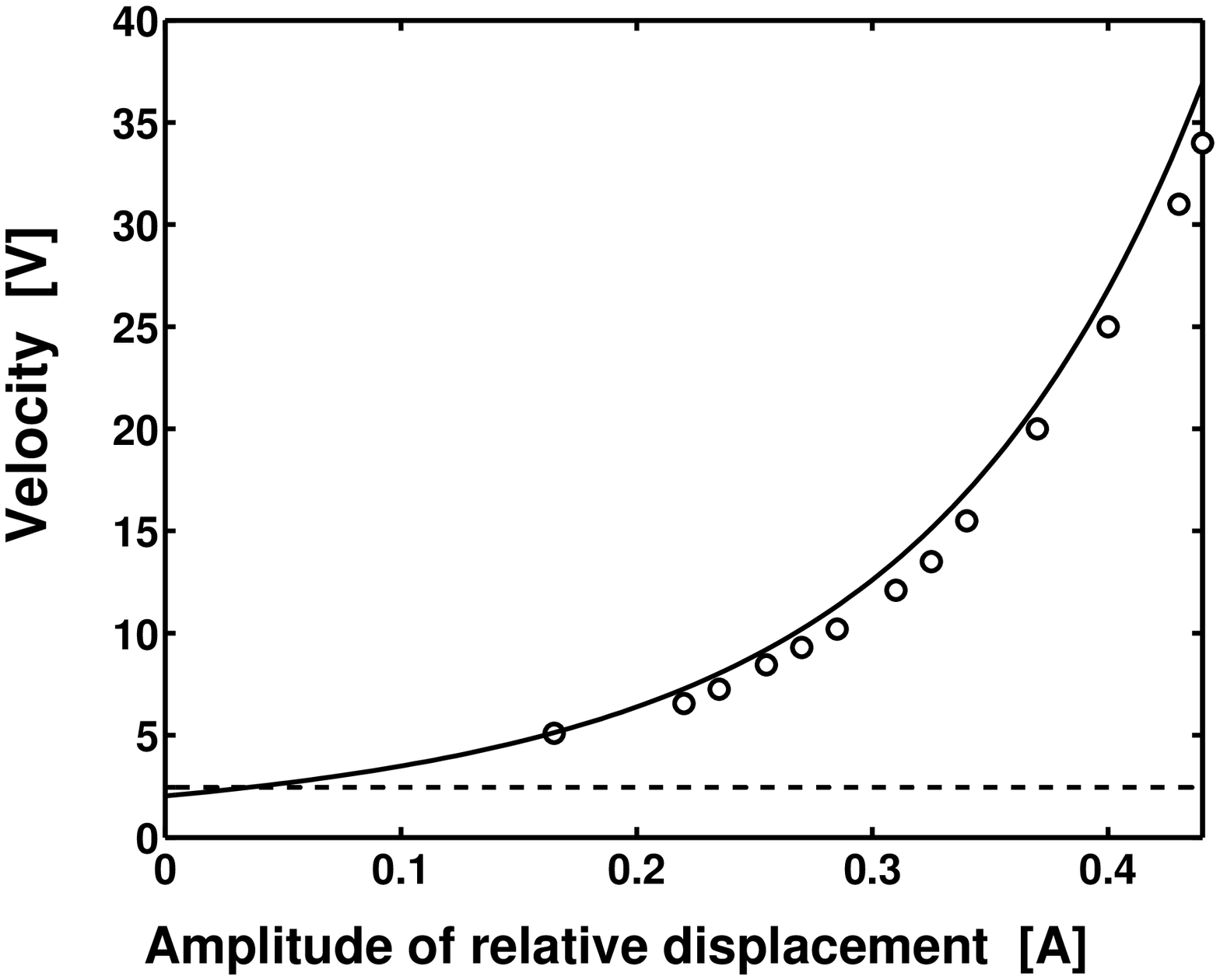}
\includegraphics[height=12cm]{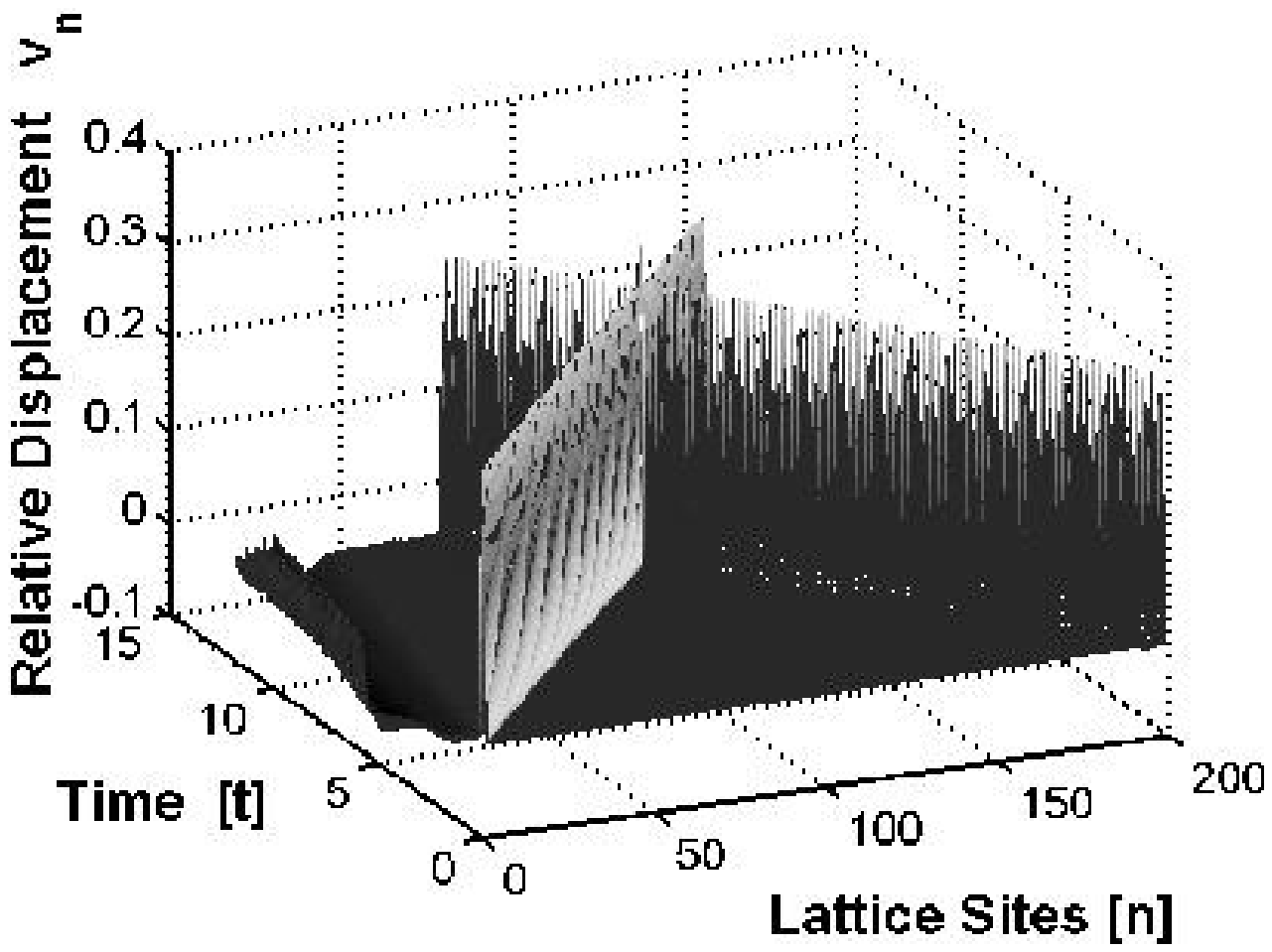}
\end{center}
\caption{\label{fig33} Left Graph: ``Magic" wavenumber ansatz for the
discrete kink-soliton velocity 
versus amplitude for a LJ interparticle potential. The analytical result 
(solid line) compares extremely well with numerics (open circles) . 
The dashed horizontal line indicates the maximal group velocity of linear waves 
(equal to $\sqrt{6}$). 
Right graph: Interaction of the DK's created at the opposite ends of the 
chain.
The strongly localized DK's retain their shapes after interaction.}
\end{figure}
%%%%%%%%%%%%%%%%%%%%%%%%%%%%%%%%%%%%%%%%%%%%

\section{Conclusions}

Summarizing, we have used the sharp pulse method to generate large amplitude, 
supersonic discrete kink-solitons with amplitude independent short localization 
length. 
These objects reduce to the well known kink-soliton solutions of the mKdV 
equation for small 
amplitudes. We have shown that large amplitude kink-solitons 
are characterized by a sinusoidal pattern with the ``magic" wavenumber $k=2\pi/3$,
and propagate with supersonic velocities increasing with the amplitude. 
Both the FPU and the more realistic LJ interparticle potential have 
been considered. The agreement between the lattice kink theory and the numerical 
experiments for both types of anharmonic lattices is so good that it encourages 
us to draw 
the general conclusion that strongly localized discrete kink-solitons 
are common for systems with two-body interparticle potentials.

\acknowledgments We thank O.M. Braun for sending us his papers on the 
Frenkel-Kontorova model 
and P. Rosenau for useful correspondence and for sending us 
his paper before publication.
This work was supported by the NATO expert visit award No PST.EV.979337 and the 
LOCNET European Commission RTN HPRN-CT-1999-00163.R.Kh. is supported by USA CRDF award 
No GP2-2311-TB-02 and NATO reintegration grant No FEL.RIG.980767.

\end{document}